\documentclass[12pt,a4paper]{article}
\pdfoutput=1 
\usepackage[T1]{fontenc}
\usepackage{rotating}
\usepackage{booktabs}
\usepackage{amsmath}
\usepackage{amssymb}
\usepackage{graphicx}
\usepackage{esint}

\usepackage[utf8x]{inputenc}
\usepackage{ucs}
\usepackage{amsmath}
\usepackage{amsfonts}
\usepackage{amssymb}

\usepackage{tensor}
\usepackage{bbm}
\usepackage{tikz} 
\newcommand{\D}{\mathcal D\!}
\DeclareMathOperator{\ad}{ad}

\def\a{\alpha}

\def\m{\mu}
\def\n{\nu}

\def\be{\begin{equation}}
\def\ee{\end{equation}}
\def\bea{\begin{eqnarray}}
\def\eea{\end{eqnarray}}
\def\ba{\begin{array}}
\def\ea{\end{array}}

\def\nn{\nonumber}

\makeatletter

\usepackage{jheppub}

\usepackage{etoolbox}
    
    \patchcmd{\maketitle}{\@fpheader}{}{}{}

\usepackage{amsfonts}

\usepackage{bbm}

\title{\boldmath Kac-Moody and Borcherds Symmetries of Six-Dimensional Chiral Supergravity}

\author[a,b]{Marc Henneaux,}
\author[a]{Victor Lekeu}

\affiliation[a]{Universit\'{e} Libre de Bruxelles and International Solvay Institutes,
ULB Campus Plaine C.P.231, B-1050 Bruxelles, Belgium.}
\affiliation[b]{Centro de Estudios Cient\'{i}ficos (CECs), Casilla 1469, Valdivia,
Chile.}

\emailAdd{henneaux@ulb.ac.be}
\emailAdd{vlekeu@ulb.ac.be}

\abstract{We investigate the conjectured infinite-dimensional hidden symmetries of six-dimensional chiral supergravity coupled to two vector multiplets and two tensor multiplets, which is known to possess the $F_{4,4}$ symmetry upon dimensional reduction to three spacetime dimensions. Two things are done.  (i) First, we analyze the geodesic equations on the coset space $F_{4,4}^{++}/K(F_{4,4}^{++})$ using the level decomposition associated with the subalgebra $\mathfrak{gl}(5)\oplus \mathfrak{sl}(2)$ of $F_{4,4}^{++}$  and show their equivalence with the bosonic equations of motion of six-dimensional chiral supergravity up to the level where the dual graviton appears.  In particular, the self-duality condition on the chiral $2$-form is automatically implemented in the sense that no dual potential appears for that $2$-form, in contradistinction with what occurs for the non chiral $p$-forms.  (ii) Second, we describe the $p$-form hierarchy of the model in terms of its $V$-duality Borcherds superalgebra, of which we compute the Cartan matrix.
}

\makeatother

\begin{document}
\maketitle \flushbottom

\section{Introduction\label{sec:Introduction}}
\setcounter{equation}{0}

Hidden symmetries of gravitational theories constitute a fascinating topic that finds its roots in the remarkable papers  \cite{Ehlers0,Ehlers1,Geroch:1970nt,Geroch2,Cremmer:1979up} uncovering unanticipated symmetry groups much bigger than the expected ones upon dimensional reduction. Following this pioneering work, it was conjectured that infinite-dimensional algebras of Kac-Moody type played a central role in the description of the symmetries of gravitational theories \cite{Julia:1980gr,Julia1,Julia:1982gx,Nicolai:1991kx,Julia:1997cy}.  The conjecture received an enormous support through the work \cite{Julia:1997cy,West:2001as,Schnakenburg:2001he,Englert:2003zs,Kleinschmidt:2003mf,West:2004st} connecting maximal supergravities to non-linear realizations of $E_{11}$,  and the work \cite{Damour:2000hv,Damour:2001sa} that reformulated the BKL oscillatory behaviour near a spacelike singularity of gravitational models \cite{Belinsky:1970ew,Belinsky:1982pk,Demaret:1986su,Damour:2000wm} as a billiard motion in the Weyl chamber of a hyperbolic Kac-Moody algebra, paving the way to a different non-linear realization of the symmetry where space and time are on distinct footings \cite{Damour:2002cu} (for reviews, see \cite{Damour:2002et,Henneaux:2007ej})\footnote{New insight on the hidden symmetries that appear in $D=4$ and $D=3$ spacetime dimensions has been derived recently in the light cone gauge \cite{Brink:2008qc,Brink:2008hv}.  It would be interesting to explore how that approach can also provide insight on the conjectured infinite-dimensional symmetry.}.  

Non-linear realizations of the infinite-dimensional conjectured symmetry spectacularly encode the correct field content of the corresponding supergravity theories, as well as the correct Chern-Simons couplings.
 In spite of these intriguing successes, however, it is fair to say that the full conjecture remains far from being proven.  Since one may view the conjectured hidden symmetries as generalizations of $p$-form dualities and gravitational duality, one may argue that one difficulty lies in the poor understanding of duality, and, in particular, in how to make it manifest from the outset. 
 
With this aspect of the problem in mind, we investigate here chiral supergravity in six dimensions coupled to abelian vector multiplets and tensor multiplets \cite{Romans:1986er,Nishino:1986dc,Sagnotti:1992qw,Nishino:1997ff,Ferrara:1997gh,Riccioni:1998th,Riccioni:1999xq}. These models involve chiral $2$-forms in a crucial way so that duality is an essential ingredient from the very beginning.  

The symmetry that appears upon dimensional reduction to 3 dimensions of pure  chiral supergravity is $B_3$  and becomes $F_4$ if one couples two abelian vector multiplets and two tensor multiplets \cite{Cremmer:1999du}.  [Only the split real forms appear here, i.e., $B_3 \equiv B_{3,3} \equiv so(4,3)$ and $F_4 \equiv F_{4,4}$.] It is this enlarged version of the theory that we shall explore.  It is natural to conjecture that the hidden symmetry underlying this model is $F_4^{++}$ (or $F_4^{+++}$).  A  preliminary billiard analysis indicates indeed that the relevant billiard table is the Weyl chamber of $F_4^{++}$ \cite{Damour:2002fz}.  The fact that it is the exceptional Lie algebra $F_4$ and its extensions that are the underlying algebras provides one further motivation in itself for investigating this model, since the hidden symmetries based on the other exceptional algebras ($E_8$ and the $E$-series, and $G_2$ \cite{G_2}) have been already analyzed, while the analysis of the dynamics based on the overextension  $F_4^{++}$ has not been made yet.

One interesting feature of six-dimensional chiral supergravity is that the actions that correctly describe the pure or extended models, i.e., such that the (anti-)self duality conditions are manifestly built-in without having to be imposed externally by hand,  are non-manifestly spacetime covariant \cite{Marcus:1982yu}. They contain as essential ingredient the actions for chiral or anti-chiral $2$-forms of \cite{Henneaux:1988gg}. These actions, although covariant, are not manifestly so, illustrating the general tension that exists between manifest spacetime covariance and manifest duality\footnote{Spacetime covariance can be made manifest by introducing gauge and auxiliary fields that appear non polynomially in the action as achieved in the interesting work \cite{Tonin,Tonin2,Pasti:1996vs,Dall'Agata:1997db}.
One may take the point of view, however, that there is a message to be learned from the tension between manifest spacetime covariance and manifest duality invariance, and that duality might be more fundamental \cite{Bunster:2012hm}.}.

It was shown in \cite{Kleinschmidt:2003mf,Kleinschmidt:2003jf,Riccioni:2007hm} that the nonlinear realization of $F_4^{+++}$ reproduces correctly the field content of chiral supergravity.  The central goal of our paper is to study more explicitly the dynamics. Our aim is to derive the geodesic motion on the coset space $F_4^{++}/K(F_4^{++})$ and to compare it with the solutions of the supergravity equations in a level expansion similar to that introduced in \cite{Damour:2002cu}, which turns out to be here a bi-level.  Here, $K(F_4^{++})$ is the ``maximal compact subalgebra" of $F_4^{++}$. It contains the maximal compact subalgebra $sp(3) \oplus su(2)$ of $F_4$. Our main result is to establish complete agreement between the two models up to (but not including) the level that involves the dual graviton. In particular, we find that the nonlinear sigma model encodes the self-duality of the chiral two-form.  

The model resembles in many respects the geodesic motion for type IIB supergravity \cite{Kleinschmidt:2004rg}, where there is a chiral 4-form, the self-duality condition of which is properly incorporated in the sigma model formulation. The $p$-form content is, however, different since $F_4$ involves not just $p$-forms of even degree, but also $p$-forms of odd degree.  The difference in the $p$-form content is best described by comparing the ``$V$-duality" Borcherds superalgebras that control the respective $p$-form algebras \cite{HenryLabordere:2002dk,Henneaux:2010ys}. We compute the $V$-duality Borcherds superalgebra for the $F_4$ model and compare and contrast its Cartan matrix with that relevant to type IIB.

Our paper is organized as follows.  In the next section (Section {\bf \ref{ChiralSugra}}) we recall the Lagrangian of the bosonic sector of six-dimensional chiral supergravity coupled to two vector multiplets and two tensor multiplets and write the equations of motion.  We also provide explicitly the Lagrangian for which the chirality condition is automatically implemented and does not have to be imposed by hand from the outside.  We study in Section {\bf \ref{Sec:F4}} the hyperbolic Kac-Moody algebra $F_4^{++}$ and give its low level roots in the decomposition with respect to the subalgebra $\mathfrak{gl}(5)\oplus \mathfrak{sl}(2)$.  We then turn in Section {\bf \ref{Sec:Sigma}} to the sigma model formulation, for which we write the equations of motion. In  Section {\bf \ref{Sec:Comparison}}, we compare the supergravity equations with those of the sigma model and provide the dictionary that make these equations match up to the level of the dual graviton.  In particular, we show how the self-duality condition on the chiral 2-form is incorporated within the sigma model. We also comment on the standard difficulties that appear at and above the dual graviton level.  In Section {\bf \ref{Sec:Borcherds}}, we determine the Borcherds structure of the $p$-form $V$-duality algebra.  In that analysis, we follow the method of  \cite{Kleinschmidt:2013em} to eliminate some ambiguities, which requires the determination of the V-duality algebras for the dimensionally reduced models in spacetime dimensions lower than 6. Section {\bf \ref{sec:Concluding-remarks}} is devoted to the conclusions where further comments on manifest duality symmetry are provided. We compare, in particular, the ways in which the self-duality condition appears in the sigma model approach and in the field theoretical description. 

While the level decomposition of $F_4^{++}$ has been studied previously in the $F_4^{+++}$ context \cite{Kleinschmidt:2003mf}, the explicit matching of the geodesic equations on the coset space $F_4^{++}/K(F_4^{++})$ with the bosonic field equations for six-dimensional chiral supergravity coupled to two vector multiplets and two tensor multiplets constitutes to our knowledge a new result, together with the determination of the corresponding Borcherds structure.  Furthermore, the self-contained Lagrangian for six-dimensional chiral supergravity, in which the self-duality condition appears as an equation of motion, has not  been written before as far as we know.

\section{Chiral supergravity} 
\setcounter{equation}{0}
\label{ChiralSugra}

\subsection{Lagrangian (standard formulation)}
The bosonic field content of $D=6$, $\mathcal{N} = (1,0)$ supergravity coupled to two tensor multiplets and two vector multiplets consists of the metric $g$, two scalars $\phi$ (dilaton) and $\psi$ (axion), two vectors $A^{\pm}$ and two $2$-forms $B$ and $C$.  The field strengths are given by
\begin{subequations}
\begin{align}
&F^+ = dA^+ + \frac{1}{\sqrt 2} \chi dA^- , \; \; 
F^- = dA^- \\
&H = dB + \frac{1}{2} A^- \wedge dA^- , \; \; G = dC - \frac{1}{\sqrt 2} \chi H - \frac{1}{2} A^+ \wedge dA^- \label{DefOfG}
\end{align}
\end{subequations}
and the Lagrangian reads explicitly:
\bea
\mathcal{L} &= & R \star \mathbbm{1} - \star d\phi \wedge d\phi - \frac{1}{2}e^{2\phi}\star d\chi \wedge d\chi -\frac{1}{2} e^{\phi} \star F^{+} \wedge F^{+} -\frac{1}{2} e^{-\phi} \star F^{-} \wedge F^{-}  \nn \\
&& - \frac{1}{2} e^{-2\phi} \star H \wedge H - \frac{1}{2} \star G \wedge G  \nn \\
&& - \frac{1}{\sqrt{2}} \chi H \wedge G - \frac{1}{2} A^{+} \wedge F^{+} \wedge H - \frac{1}{2} A^{+} \wedge F^{-} \wedge G, \label{L0for6Sugra}
\eea
The self-duality condition $\star G = G$ must be imposed after varying $\mathcal L$ to get the equations of motion.  As shown in \cite{Cremmer:1999du}, the $D=6$ Lagrangian (\ref{L0for6Sugra}) can be viewed as the oxidation endpoint of the theory with $F_4$ symmetry in 3 spacetime dimensions, i.e., $D=3$ gravity coupled to the nonlinear sigma model $F_4/(Sp(3) \times SU(2))$.

\subsection{Equations of motion}

Extremizing the action with respect to the metric yields the Einstein equations:
\be R_{\mu\nu} - \frac{1}{2} g_{\mu\nu} R = T_{\mu\nu} \ee
where  the energy-momentum tensor reads

\begin{align}
T_{\mu\nu} &= \partial_{\mu}\phi \partial_{\nu}\phi + \frac{1}{2}e^{2\phi}\partial_{\mu}\chi \partial_{\nu}\chi +\frac{1}{2} e^{\phi} F^{+}_{\mu\rho} F\indices{^+_\nu^\rho} + \frac{1}{2} e^{-\phi} F^{-}_{\mu\rho} F\indices{^-_\nu^\rho}  \nn \\
&\quad +\frac{1}{4}e^{-2\phi} H_{\mu\rho\sigma}H\indices{_\nu^\rho^\sigma}+\frac{1}{4}G_{\mu\rho\sigma}G\indices{_\nu^\rho^\sigma} \nn \\
&\quad -\frac{1}{2}g_{\mu\nu} \left[ \partial_{\rho}\phi \partial^{\rho}\phi + \frac{1}{2}e^{2\phi}\partial_{\rho}\chi \partial^{\rho}\chi + \frac{1}{4} e^{\phi} F^{+}_{\rho\sigma} F^{+{\rho\sigma}} + \frac{1}{4} e^{-\phi} F^{-}_{\rho\sigma} F^{-{\rho\sigma}}\right. \nn\\
&\qquad\qquad\quad \left.+\frac{1}{12}e^{-2\phi} H_{\rho\sigma\lambda}H^{\rho\sigma\lambda}+\frac{1}{12} G_{\rho\sigma\lambda}G^{\rho\sigma\lambda} \right].
\end{align}
Since $D=6$, an equivalent form of the Einstein equations is \[ R_{\mu\nu} = T_{\mu\nu} - \frac{1}{4} g_{\mu\nu} T\indices{_\rho^\rho}. \]

For the other fields, one has the Bianchi identities
\begin{subequations}
\begin{align}
&d d \chi = d d\phi = 0 \\
&dF^+ = \frac{1}{\sqrt 2} d\chi \wedge F^- , \; \; dF^- = 0 \\
&dH = \frac{1}{2} F^- \wedge F^- , \; \; dG = - \frac{1}{\sqrt 2} d\chi \wedge H - \frac{1}{2} F^+ \wedge F^-
\end{align}
\end{subequations}
and the equations of motion
\begin{subequations}
\begin{align}
& d \star G = -\frac{1}{\sqrt{2}} d\chi \wedge H -\frac{1}{2} F^{+} \wedge F^{-} , \; \;  d \left( e^{-2\phi} \star H \right) = \sqrt{2} d\chi \wedge G - \frac{1}{2} F^{+} \wedge F^{+}  \\
&d \left( e^\phi \star F^+ \right) = - G \wedge F^- - H \wedge F^+ \\
& d \left( e^{-\phi} \star F^- \right) = e^{-2\phi} \star H \wedge F^- - G\wedge F^+ - \frac{1}{\sqrt 2} e^\phi d \chi \wedge \star F^+ \\
& d \left( e^{2\phi} \star d\chi \right) = -\frac{1}{\sqrt 2} e^\phi \star F^+ \wedge F^- + \sqrt{2} G \wedge H \\
& d \star d\phi = - \frac{1}{2}e^{2\phi}\star d\chi \wedge d\chi -\frac{1}{4} e^{\phi} \star F^{+} \wedge F^{+} + \frac{1}{4} e^{-\phi} \star F^{-} \wedge F^{-} + \frac{1}{2} e^{-2\phi} \star H \wedge H
\end{align}
\end{subequations}
Note that the Bianchi identity for $G$ and its equation of motion indeed consistently allow for $\star G = G$.

Taking the Hodge dual of these equations and expressing them in components, we get for the Bianchi identities
\begin{subequations}
\begin{align}
 \partial_\mu\left(\sqrt{-g}\,(\star d\phi)\indices{^{\mu \nu\rho\sigma\tau}}\right) &= 0 \\
 \partial_\mu\left(\sqrt{-g}\,(\star d\chi)\indices{^{\mu \nu\rho\sigma\tau}}\right) &= 0 \\
\partial_\mu\left(\sqrt{-g}\,(\star F^-)^{\mu \nu\rho\sigma}\right) &= 0 \\
 \partial_\mu\left(\sqrt{-g}\,(\star F^+)^{\mu \nu\rho\sigma}\right) &= - \frac{1}{2\sqrt 2}  \,\varepsilon\indices{^{\nu\rho\sigma\alpha\beta\gamma}} \partial_\alpha \chi F_{-\beta\gamma} \\
 \partial_\mu\left(\sqrt{-g}\,(\star H)\indices{^{\mu \nu\rho}}\right) &= \frac{1}{8}  \,\varepsilon\indices{^{\nu\rho\alpha\beta\gamma\delta}} F_{-\alpha\beta} F_{-\gamma \delta} \\
 \partial_\mu\left(\sqrt{-g}\,(\star G)\indices{^{\mu \nu\rho}}\right) &=  \varepsilon\indices{^{\nu\rho\alpha\beta\gamma\delta}}\left( -\frac{1}{6\sqrt{2}} \partial_\alpha \chi H_{\beta\gamma\delta} -\frac{1}{8} F_{+\alpha\beta} F_{-\gamma\delta} \right)
\end{align}
\end{subequations}
and for the equations of motion
\begin{subequations}
\begin{align}
 \partial_\mu\left(\sqrt{-g}\,G\indices{^{\mu \nu\rho}}\right) &=  \varepsilon\indices{^{\nu\rho\alpha\beta\gamma\delta}}\left( -\frac{1}{6\sqrt{2}} \partial_\alpha \chi H_{\beta\gamma\delta} -\frac{1}{8} F_{+\alpha\beta} F_{-\gamma\delta} \right) \\
\frac{1}{\sqrt{-g}} \partial_\mu\left(\sqrt{-g}\,e^{-2\phi}H\indices{^{\mu \nu\rho}}\right) &= \sqrt{2}\, G^{\nu\rho\alpha} \partial_\alpha \chi - \frac{1}{8  \sqrt{-g}} \,\varepsilon\indices{^{\nu\rho\alpha\beta\gamma\delta}} F_{+\alpha\beta} F_{+\gamma\delta} \\
\frac{1}{\sqrt{-g}} \partial_\mu\left(\sqrt{-g}\,e^{\phi}F\indices{^{+\mu \nu}}\right) &= -\frac{1}{2} G^{\nu\alpha\beta}F_{-\alpha\beta} - \frac{1}{12  \sqrt{-g}} \,\varepsilon\indices{^{\nu\alpha\beta\gamma\delta\epsilon}} H_{\alpha\beta\gamma}F_{+\delta\epsilon} \\
\frac{1}{\sqrt{-g}} \partial_\mu\left(\sqrt{-g}\,e^{-\phi}F\indices{^{-\mu \nu}}\right) &= \frac{1}{2} e^{-2\phi} H^{\nu\alpha\beta} F_{-\alpha\beta} + \frac{1}{\sqrt 2} e^\phi F^{+\nu\alpha} \partial_\alpha \chi  - \frac{1}{2} G^{\nu\alpha\beta}F^+_{\alpha\beta} \\
\frac{1}{\sqrt{-g}}\partial_\mu \left( \sqrt{-g}\,e^{2\phi} \partial^\mu \chi \right) &= \frac{1}{2\sqrt 2} e^{\phi} F^+_{\alpha\beta} F^{-\alpha\beta} - \frac{1}{3\sqrt 2} G_{\alpha\beta\gamma}H^{\alpha\beta\gamma} \\
\frac{1}{\sqrt{-g}}\partial_\mu \!\left( \sqrt{-g} \,\partial^\mu \phi \right) &= \frac{1}{2} e^{2\phi} \partial_\alpha \chi \partial^\alpha \chi + \frac{1}{8} e^{\phi} F^+_{\alpha\beta} F^{+ \alpha\beta} - \frac{1}{8} e^{-\phi} F^-_{\alpha\beta} F^{-\alpha\beta} \nn \\ 
&\quad - \frac{1}{12} e^{-2\phi} H_{\alpha\beta\gamma} H^{\alpha\beta\gamma}
\end{align}
\end{subequations}
Our conventions are
\be \varepsilon_{01\dots 5} = 1, \; \varepsilon^{01\dots 5}=-1, \; (\star \omega)_{\nu_1\nu_2\dots \nu_q} = \frac{\sqrt{-g}}{p!} \varepsilon\indices{_{\nu_1\nu_2\dots\nu_q\mu_1\mu_2\dots\mu_p}} \omega^{\mu_1\mu_2\dots\mu_p}. \ee

\subsection{Lagrangian with self-duality built in}
It is somewhat unsatisfactory to have to implement by hand the self-duality condition $\star G = G$ on the two-form $C$.  A satisfactory action principle should be self-contained.  We give such an action principle here.  It generalizes the free action of \cite{Henneaux:1988gg} by including the  Chern-Simons couplings.  The easisest way to derive it from the Lagrangian (\ref{L0for6Sugra}) is to follow the steps of  \cite{Bunster:2011qp,Bekaert:1998yp}.  

We can write the Lagrangian (\ref{L0for6Sugra}) as
\begin{align}
\mathcal{L} &= - \frac{1}{2} \star G \wedge G + D \wedge G + \mathcal{L}_0,
\end{align}
where the $3$-form $D$ is given by 
\be D=- \frac{1}{\sqrt 2} \chi H - \frac{1}{2} A^+ \wedge dA^-
\ee
(see (\ref{DefOfG})), and where $\mathcal{L}_0$ does not contain the $2$-form $C$.  We now:
\begin{itemize}
\item Go to the Hamiltonian formalism only for the $2$-form $C$, while keeping the other fields in second order form; i.e., perform the Legendre transformation only on the time derivatives $\dot{C}_{ij}$ of $C$ and the conjugate momenta $\pi^{ij}$ ,
\be
\pi^{ij} = \frac{\partial \mathcal{L}}{\partial \dot{C}_{ij}}
\ee
\item Solve the Gauss constraints $\partial_i \pi^{ij} = 0$ that follows from varying the action with respect to the Lagrange multipliers $C_{0i}$ by introducing a second 2-form potential $Z_{ij}$,
\be \pi^{ij} = \frac{1}{2} \varepsilon^{ijklm} \partial_k Z_{lm} \ee
to get an action $S[C_{ij},Z_{ij}]$ that involves the two spatial $2$-form potentials $C_{ij}$ and $Z_{ij}$ (plus $\int d^6x \mathcal{L}_0$ which remains unaffected by all these steps).
\item Make the change of variables
\begin{subequations}
\begin{align}
C_{ij} &= \sqrt{2}(C^+_{ij} - C^-_{ij}) \\
Z_{ij} &= \sqrt{2}(C^+_{ij} + C^-_{ij}).
\end{align}
\end{subequations}
Under this change of variables, the action  splits as a sum of an interacting action for the chiral part $C^+_{ij}$ and a {\em free} action for the anti-chiral part $C^-_{ij}$.  The free action for the anti-chiral part $C^-_{ij}$ can be consistently dropped, leaving one with the action describing correctly the interacting chiral $2$-form without superfluous degrees of freedom.
\end{itemize}

If one follows this procedure, one gets the action
\begin{align}
S = \int d^6x &\left[ \frac{1}{2} \varepsilon^{ijklm} \left( (\partial_0 C^+_{ij} + \frac{1}{\sqrt{2}} D_{0ij}) \partial_{k} C^+_{lm} - N^p G^+_{pij} G^+_{klm}\right) \right.  \nn \\
&\left. - \frac{1}{3} N \sqrt{\mathrm{g}} \mathrm{g}^{ip}\mathrm{g}^{jq}\mathrm{g}^{kr} G^+_{ijk}G^+_{pqr}   + \frac{1}{6\sqrt 2} \varepsilon^{ijklm} D_{0ij} G^+_{klm} + \mathcal{L}_0 \right], \label{ChiralAction0}
\end{align}
where
\be G^+_{mnr} = 3 \partial_{[m} C^+_{nr]} + \frac{1}{\sqrt 2} D_{mnr} \ee
and where
$N$ is the lapse, $N^k$ the shift, $\mathrm{g}_{ij}$ the spatial metric, and the convention for the spatial $\varepsilon$ tensor is $\varepsilon^{12345}=1$.

It is useful, in order to keep track of the gauge symmetries, to introduce the time components $C^+_{0j}$ in the kinetic term of (\ref{ChiralAction0}) so as to make the invariant field strength $G^+_{0ij}$ appear.  This can be done at no cost because these extra components $C^+_{0j}$ drop out of the action by integration by parts.
One gets
\begin{align}
S = \int d^6x &\left[ \frac{1}{6} \varepsilon^{ijklm} ( G^+_{0ij} G^+_{klm} - N^p G^+_{pij} G^+_{klm}) - \frac{1}{3} N \sqrt{\mathrm{g}} \mathrm{g}^{ip}\mathrm{g}^{jq}\mathrm{g}^{kr} G^+_{ijk}G^+_{pqr}  \right.  \nn \\
& \left. + \frac{1}{6\sqrt 2} \varepsilon^{ijklm} D_{0ij} G^+_{klm} - \frac{1}{6\sqrt 2} \varepsilon^{ijklm} G^+_{0ij} D_{klm} + \mathcal{L}_0 \right], \label{ChiralAction1}
\end{align}
where
\be G^+_{\mu \nu \rho} = 3 \partial_{[\mu} C^+_{\nu\rho]} + \frac{1}{\sqrt 2} D_{\mu \nu \rho} . \ee

Restoring the expressions for $D$ and $\mathcal{L}_0$, and noticing that the expression
$$ \int d^6x \left[ \frac{1}{6\sqrt 2} \varepsilon^{ijklm} D_{0ij} G^+_{klm} - \frac{1}{6\sqrt 2} \varepsilon^{ijklm} G^+_{0ij} D_{klm} \right]$$
 is equal to  $-\int d^6x \frac{\sqrt 2}{6^2} \varepsilon^{\mu\nu\rho\sigma\lambda\tau}D_{\mu\nu\rho}G^+_{\sigma\lambda\tau} = \int d^6x \sqrt 2 D \wedge G^+$, 
the final form of the action is found to be
\begin{align}
S = \int d^6x &\left[ R \star \mathbbm{1} - \star d\phi \wedge d\phi - \frac{1}{2}e^{2\phi}\star d\chi \wedge d\chi -\frac{1}{2} e^{\phi} \star F^{+} \wedge F^{+} -\frac{1}{2} e^{-\phi} \star F^{-} \wedge F^{-} \right. \nn \\
& + \frac{1}{6} \varepsilon^{ijklm} ( G^+_{0ij} G^+_{klm} - N^p G^+_{pij} G^+_{klm}) - \frac{1}{3} N \sqrt{\mathrm{g}} \mathrm{g}^{ip}\mathrm{g}^{jq}\mathrm{g}^{kr} G^+_{ijk}G^+_{pqr} \nn \\
&\left. - \frac{1}{2} e^{-2\phi} \star H \wedge H - \chi H \wedge G^+ - \frac{1}{2} A^{+} \wedge F^{+} \wedge H - \frac{1}{\sqrt 2} A^{+} \wedge F^{-} \wedge G^+ \right], \label{ChiralAction2}
\end{align}
where
\be G^+ = dC^+ - \frac{1}{2} \chi H - \frac{1}{2\sqrt 2} A^+ \wedge dA^- . \ee
This action has the gauge symmetries:
\begin{subequations}
\begin{align}
\delta A^+ & = d \epsilon^+, \; \; \delta A^- = d \epsilon^- \\
\delta B &= d \Lambda - \frac12 \epsilon^- dA^- \\
\delta C^+  & = d \Xi + \frac{1}{2 \sqrt{2}} \epsilon^+ dA^- \label{CXi}
\end{align}
\end{subequations}
under which the field strengths are invariant.  Here $\epsilon^+$ and $\epsilon^-$ are $0$-forms, while $\Lambda$ and $\Xi$ are $1$-forms.  In addition to (\ref{CXi}), the action is also invariant under arbitrary shifts of $C^+_{0i}$ which occurs only through a total derivative,
\be
\delta C^+_{0i} = \Psi_i \label{CPsi}.
\ee
The gauge symmetries (\ref{CXi}) and (\ref{CPsi}) are of course not independent.
 
 Contrary to the original action (\ref{L0for6Sugra}), the action (\ref{ChiralAction2}) carries no superfluous degrees of freedom that have to be eliminated by hand. It correctly describes, in a self-contained manner,  the coupling of a chiral $2$-form with the other degrees of freedom of six-dimensional chiral supergravity.
It is the analog of the action of \cite{Bekaert:1999sq} for type IIB supergravity in ten dimensions.

\section{Level decomposition of $F_4^{++}$} 
\setcounter{equation}{0}
\label{Sec:F4}

\subsection{Dynkin diagram and Cartan matrix}

The Dynkin diagram of $F_4^{++}$ is:
\begin{center}
\begin{tikzpicture}[scale=1]
\tikzset{v/.style={circle,fill,inner sep=0pt,minimum size=3.5pt,draw}}
\draw (0,0) node[v]{} node[below]{$\alpha_1$} -- (1,0) node[v]{} node[below]{$\alpha_2$} -- (2,0) node[v]{} node[below]{$\alpha_3$}{} -- (3,0) node[v]{} node[below]{$\alpha_4$}{};
\draw (4,0) node[v]{} node[below]{$\alpha_5$}{} -- (5,0) node[v]{} node[below]{$\alpha_6$}{};
\draw (3, 0.05) -- (4, 0.05);
\draw (3, -0.05) -- (4, -0.05);
\draw (3.4, 0.15) -- (3.6, 0) -- (3.4, -0.15);
\end{tikzpicture}
\end{center}
corresponding to the
Cartan matrix:
\be A=\begin{pmatrix}
2 & -1 & 0 & 0 & 0 & 0 \\ 
-1 & 2 & -1 & 0 & 0 & 0 \\ 
0 & -1 & 2 & -1 & 0 & 0 \\ 
0 & 0 & -1 & 2 & -1 & 0 \\ 
0 & 0 & 0 & -2 & 2 & -1 \\ 
0 & 0 & 0 & 0 & -1 & 2
\end{pmatrix} \ee
We normalize the long real roots to have length squared equal to $2$, e.g. $(\alpha_1 \vert \alpha_1) = 2$.

\subsection{$\mathfrak{gl}(5)$-subalgebra}

The first four roots of $F_4^{++}$ define an $A_4$-subalgebra with Chevalley generators \\ $\{ h_i, e_i, f_i | i = 1,2,3,4 \}$, which can be enlarged to a $\mathfrak{gl}(5)$-subalgebra by adding an appropriate combination of the Cartan generators $h_5$ and $h_6$ as follows.

The usual presentation of $\mathfrak{gl}(5)$ is given in terms of the generators  $K\indices{^a_b}$, where $a$ and $b$ go from $1$ to $5$, which satisfy the commutation relations
\be [K\indices{^a_b},K\indices{^c_d}] = \delta^c_b K\indices{^a_d} - \delta^a_d K\indices{^c_b} . \label{gl5} \ee
The invariant bilinear form is
\be (K\indices{^a_b} | K\indices{^c_d}) = \delta^a_d \delta^c_b - \delta^a_b \delta^c_d. \ee
The explicit embedding of $\mathfrak{gl}(5)$ in $F_4^{++}$ is given by
\begin{subequations}
\begin{align}
e_i &= K\indices{^i_{i+1}} \\
f_i &= K\indices{^{i+1}_i} \\
h_i &= K\indices{^i_i} - K\indices{^{i+1}_{i+1}}
\end{align}
\end{subequations}
for $i=1,2,3,4$ (this gives the embedding of $\mathfrak{sl}(5)$) and
\be h_5 + \frac{1}{2} h_6 = -\frac{1}{2}( K\indices{^1_1} + K\indices{^2_2} + K\indices{^3_3} + K\indices{^4_4} ) + \frac{3}{2} K\indices{^5_5} \ee
or conversely,
\be \sum_{a=1}^5 K\indices{^a_a} = - 4 h_1 - 8 h_2 - 12 h_3 - 16 h_4 - 10 h_5 - 5 h_6. \ee

One can take as basis of the Cartan subalgebra of $F_4^{++}$ the five $K\indices{^a_a}$ and one additional independent Cartan generator, which we choose to be $h_6=K$.

\subsection{Definition of level}

The algebra $F_4^{++}$ can be decomposed in terms of irreducible representations of this subalgebra.    We define  the (bi-valued) level  $(l,l')$ of a Cartan element to be $(0,0)$, and that of a root vector $e_\a$ associated with the root $\a$ by the formula
\begin{equation} \alpha = \sum_{i=1}^4 m_i \alpha_i + l \alpha_5 + l' \alpha_6 . \end{equation}
Subspaces of $F_4^{++}$ corresponding to definite values of $(l,l')$ are invariant subspaces under the action of $\mathfrak{gl}(5)$ and decompose under irreducible representations of $\mathfrak{gl}(5)$.  In fact, subspaces corresponding to a definite value of $l$ (with $l'$ unspecified) form representations of $\mathfrak{gl}(5) \oplus \mathfrak{sl}(2)$, where $\mathfrak{sl}(2)$ is the subalgebra associated with the last node, generated by $\{h_6, e_6, f_6 \}$.  It will convenient, however, to fix both $l$ and $l'$ to begin with, and consider how different representations of $\mathfrak{gl}(5)$ with same $l$ and different $l'$ combine to form representations of $\mathfrak{sl}(2)$ only later.

At level $(l,l') = (0,0)$, we have the $\mathfrak{gl}(5)$-subalgebra with Chevalley generators $\{h_i, e_i, f_i | i = 1,2,3,4 \}$, along with the extra Cartan generators $h_6$. 

\subsection{Low level decomposition of $F_4^{++}$}
Since our  goal is to describe the coset space $F_4^{++}/K(F_4^{++})$, we shall focus on positive roots for which
$$ m_i \geq 0, \; \; \; l\geq 0, \; \; \; l' \geq 0.$$
The negative part of the algebra can be obtained by using the Chevalley involution.
In order to analyse the $\mathfrak{gl}(5)$-representation content of $F_4^{++}$,  we follow the method of  \cite{Damour:2002cu,Nicolai:2003fw}. The fundamental weights of $A_4$ are defined by
\be (\mu_i \vert \alpha_j) = \delta_{ij} \; \; \; \; i, j = 1,2,3,4 \ee
(with $\mu_i$ in the linear span of the $\alpha_j$'s).  Explicitly,
\be
\m_i = \sum_j \a_j B_{ji}
\ee 
where the symmetric matrix $(B_{ij})$ is the inverse matrix to the (symmetric) Cartan matrix of $A_4$,
\be
(B_{ij} ) = \frac15
\begin{pmatrix} 4 & 3 & 2 & 1 \\
3 & 6 & 4 & 2 \\
2 & 4 & 6 & 3 \\
1 & 2 & 3 & 4 
\end{pmatrix} .
\ee
The scalar products of the fundamental weights  $(\m_i \vert \m_j)$ are given by 
\be (\m_i \vert \m_j) =  B_{ij}. \ee

The root vector $e_\a$ associated with the positive root  $\a$  of $F_4^{++}$ is a weight vector for $A_4$, i.e.,
\be [h_i, e_\a] = \mu(h_i) e_\a, \; \; \; \; i= 1,2,3,4, \ee
where $\mu$ is a linear combination of the fundamental weights $\mu_i$.  On the other hand
\be [h_i, e_\a] = \a(h_i) e_\a \ee
which implies that $\a - \m$ is such that $(\a - \m)(h_i) = 0$, i.e., is orthogonal to the $4$-plane spanned by $\a_i$ ($i=1,2,3,4$).  If one denotes by $\n$ the unit normal to that  $4$-plane in the hyperplane spanned by $\a_i$ and $\a_5$ such that $(\n \vert \a_5) >0$, one easily finds 
$ 
\n = \sqrt{5} (\a_5 + \m_4)
$
since $(\a_5 \vert \a_5) = 1 $ and $(\m_4 \vert \a_5) = - \frac45$.  In other words,
\be \a_5 = \frac{1}{\sqrt{5}} \n - \m_4. \ee
The $A_4$-weight $\mu$ associated with the root $\a$ is thus $\a - \frac{l}{\sqrt{5}} \n - l' \a_6$ since the difference $\a - \mu = \frac{l}{\sqrt{5}} \n + l' \a_6$ is indeed orthogonal to the $4$-plane spanned by $\a_i$ ($i=1,2,3,4$).  Expanding $\mu$ in the basis of fundamental weights,
$\mu = - \sum_i p_i \mu_i$,  yields then the expression
\be m_1 \a_1 + m_2 \a_2 + m_3 \a_3 + m_4 \a_4 = - p_1 \m_1 - p_2 \m_2 - p_3 \m_3 + (l - p_4) \m_4. \ee
We thus have the relationships
\begin{subequations}
\begin{align}
m_1 &= \frac{1}{5} (l - 4 p_1 - 3 p_2 - 2 p_3 - p_4)  \\
m_2 &= \frac{1}{5} (2 l - 3 p_1 - 6 p_2 - 4 p_3 - 2 p_4)  \\
m_3 &= \frac{1}{5} (3 l - 2 p_1 - 4 p_2 - 6 p_3 - 3 p_4)  \\
m_4 &= \frac{1}{5} (4 l - p_1 - 2 p_2 - 3 p_3 - 4 p_4)  .
\end{align}
\end{subequations}

Now, among the vectors $e_\a$ transforming in a given irreducible representation of $A_4$, there is one lowest weight vector annihilated by all the $f_i$'s ($i=1,2,3,4$).   The lowest weight vectors are the easiest to identify. For instance $e_5$ is a lowest weight vector since $[f_i, e_5] = 0$.  Accordingly, we shall determine the irreducible $A_4$-representations that appear in the decomposition of the positive Borel subalgebra of $F_4^{++}$ by searching for their lowest weights. 

For a lowest weight $\mu$, the integers $p_i$'s are all non-negative and define the Dynkin coefficients of the representation (this explains why we have taken the coefficients in the expansion of $\mu$ to be $-p_i$). In terms of Young tableaux, $p_1$ is the number of columns of height $4$,  $p_2$ is the number of columns of height $3$, $p_3$ is the number of columns of height $2$ and $p_4$ is the number of columns of height $1$.

Let $\Lambda$ be a positive root of $F_4^{++}$ defining a lowest weight of an $A_4$-representation. 
The constraints that the $m_i$'s be non-negative integers and the condition $(\Lambda\vert \Lambda) \leq 2$ arising from the fact that $\Lambda$ is a root read
\begin{subequations} \label{conditions}
\begin{align}
&  \frac{1}{5} (l - 4 p_1 - 3 p_2 - 2 p_3 - p_4) \in \mathbb{N} \\
&  \frac{1}{5} (2 l - 3 p_1 - 6 p_2 - 4 p_3 - 2 p_4) \in \mathbb{N} \\
  &\frac{1}{5} (3 l - 2 p_1 - 4 p_2 - 6 p_3 - 3 p_4) \in \mathbb{N} \\
 &\frac{1}{5} (4 l - p_1 - 2 p_2 - 3 p_3 - 4 p_4) \in \mathbb{N} \\
& (\Lambda | \Lambda) = \sum_{i,j=1}^4 B_{ij} p_i p_j + \frac{1}{5} l^2 + l'^2 - ll' \leq 2 
\end{align}
\end{subequations}
with $p_i \geq 0$ and $\mathbb{N} = \{0, 1, 2 , \cdots \}$ the set of non-negative integers.

These inequalities determine the low level roots.  Solutions up to $l=4$ are easily verified to be given by:
\begin{center}
{\bf Table 1: } Low level decomposition of $F_4^{++}$

\begin{tabular}{c|c|c|c|c|c|c}\label{table:F4}
$l$ & $l'$ & $[p_1,p_2,p_3,p_4]$ & $(m_1,m_2,m_3,m_4)$ & $(\Lambda|\Lambda)$ & $F_4^{++}$ generators & $\sigma$-model field \\ \hline \hline
$0$ & $1$ & $[0,0,0,0]$ & $(0,0,0,0)$ & $1$ & $E$ & $\psi$ \\ \hline
$1$ & $0$ & $[0,0,0,1]$ & $(0,0,0,0)$ & $1$ & $E^a$ & $A_a$ \\
 & $1$ & $[0,0,0,1]$ & $(0,0,0,0)$ & $1$ & $E'^a=[E^a,E]$ & $B_a$ \\ \hline
$2$ & $0$ & $[0,0,1,0]$ & $(0,0,0,1)$ & $2$ & $E^{ab}=[E^a,E^b]$ & $A_{ab}$ \\
 & $1$ & $[0,0,1,0]$ & $(0,0,0,1)$ & $1$ & $E'^{ab}=[E^{ab},E]$ & $B_{ab}$ \\
 & $2$ & $[0,0,1,0]$ & $(0,0,0,1)$ & $2$ & $E''^{ab}=[E'^{ab},E]$ & $C_{ab}$ \\ \hline
$3$ & $1$ & $[0,1,0,0]$ & $(0,0,1,2)$ & $1$ & $E^{abc}=[E'^{ab},E^{c}]$ & $A_{abc}$ \\
 & $2$ & $[0,1,0,0]$ & $(0,0,1,2)$ & $1$ & $E'^{abc}=[E^{abc},E]$ & $B_{abc}$ \\ \hline
$4$ & $1$ & $[1,0,0,0]$ & $(0,1,2,3)$ & $1$ & $E^{abcd}=[E^{abc},E^{d}]$ & $A_{abcd}$ \\
& $2$ & $[0,1,0,1]$ & $(0,0,1,2)$ & $2$ & $E^{abc|d}=[E'^{\langle abc},E^{d\rangle}]$ & $A_{abc|d}$ \\
&  & $[1,0,0,0]$ & $(0,1,2,3)$ & $0$ & $E'^{abcd}=[E^{abcd},E]$ & $B_{abcd}$ \\
& $3$ & $[1,0,0,0]$ & $(0,1,2,3)$ & $1$ & $E''^{abcd}=[E'^{abcd},E]$ & $C_{abcd}$ \\ \hline
\end{tabular}
\end{center}
The brackets $\langle \dots \rangle$ indicate here projection on the Young tableau symmetry corresponding to $[0,1,0,1]$.

The conditions (\ref{conditions}) are necessary conditions for the set of integers $[p_1,p_2,p_3,p_4]$ to define a representation that appears in $F_4^{++}$.  These conditions are also sufficient here because $F_4^{++}$ is hyperbolic so that one can apply Proposition 5.10 of \cite{Kac:1990gs} to verify that the root lattice points labelled by the above $(m_i, l, l')$ are indeed roots. Since the real roots are non degenerate, the representations for which the lowest state vector $\Lambda$ has strictly positive norm occur once and only once. This is also true for the representation $[1,0,0,0]$ with $l=4$ and $l'=2$ for the following reason. The root $ \a_2 + 2 \a_3 + 3 \a_4 + 4 \a_5 + 2 \a_6$, which has $m_1=0$,  is the null root of the untwisted affine Kac-Moody algebra $F_4^+$ and is degenerate a number of times equal to the rank of $F_4$, i.e., $4$.  It occurs three times as a non lowest state vector of the representation $[0,1,0,1]$ with $l=4$ and $l' = 2$  characterized by a mixed Young tableau with one column of three boxes and one column of one box (``dual graviton").  It must therefore occur a fourth time in another representation with same values of $l$ and $l'$, which is precisely the representation $[1,0,0,0]$ with $l=4$ and $l'=2$.

As we mentioned above, the representations that differ only in the value of $l'$ combine to form representations of the  subalgebra $\mathfrak{sl}(2)$ corresponding to the last node $6$ of the Dynkin diagram.  The generator $E$ is the raising operator for those representations.   So, the representation at $l=0$ is a $\mathfrak{sl}(2)$-singlet, the representations at $l=1$ and $l=3$ are doublets, those at $l=2$ form a triplet and finally, those at $l=4$ form a triplet and a singlet. 

The decomposition of the hyperbolic algebra $F_4^{++}$ can be continued at higher levels following the procedure of \cite{Nicolai:2003fw} but this will not be needed here.  Note that Table {\bf 1} matches the level decomposition of $F_4^{+++}$ given in \cite{Kleinschmidt:2003mf,Kleinschmidt:2003jf}.

\subsection{Commutation relations}

The low-level commutation relations of $F_4^{++}$  are easy to work out. First, the commutation relations that involve the $\mathfrak{gl}(5)$ generators $K\indices{^a_b}$ are, besides (\ref{gl5}), simply given by  the usual action of $\mathfrak{gl}(5)$ on tensors, for example
\be
[K\indices{^a_b},E] = 0, \; \; [K\indices{^a_b},E^c] = \delta^c_b E^a, \; \; [K\indices{^a_b},F_{cd}] = -\delta^a_c F_{bd} - \delta^a_d F_{cb},
\ee
and so on. Here, $F_{a b \dots}=-\tau(E^{a b \dots})$ where $\tau$ is the Chevalley involution.

Second, the action of the Cartan generator $K \equiv h_6$, which is a $\mathfrak{gl}(5)$ scalar, reads
\be [K,E^{(l,l')}] = (2l'-l) E^{(l,l')} \label{actionofK} \ee
for any generator $E^{(l,l')}$ at level $(l,l')$.

Third, consider the commutation relations of the raising operators among themselves.  Some commutators are given by the defining relations in Table {\bf 1}. Some other commutators are automatically zero because there is no generator with the required symmetry at the required level, for example
\begin{align}
[E'^a,E] &= 0 \qquad\text{(no generator at level $(1,2)$),} \nn \\
[E'^{(a},E^{b)}] &= 0 \qquad\text{(no symmetric generator at level $(2,1)$),} \nn \\
[E^{ab},E^c] &= 0 \qquad\text{(no generator at level $(3,0)$).}
\end{align}
The other commutators not in that list are computed using the Jacobi identity and the above property. Up to level $(4,1)$, the nontrivial ones (=that cannot be obtained just by using antisymmetry of the commutator and antisymmetry in the indices) are
\begin{align}
& [E'^a,E^b] = \frac{1}{2} E'^{ab} , \; \; 
[E'^a,E'^b] = \frac{1}{2} E''^{ab} , \; \; 
[E^{ab},E'^c] = - E^{abc} \nn \\
& [ E'^{ab},E'^c] = - E'^{abc} , \; \; 
[E''^{ab},E^c] = 2 E'^{abc}, \; \; 
[E'^{ab},E^{cd}] = 2 E^{abcd} .
\end{align}

Similar commutators hold on the negative side of the algebra and are simply obtained by using the Chevalley involution. The last class of commutation relations involving the raising operators with the lowering operators and can also be recursively computed starting from the Chevalley relations. Here are a few examples,
\begin{align}
[E^a,F_b] &= 2K\indices{^a_b} - \frac{1}{2}\left(\sum_{e=1}^5 K\indices{^e_e} + K\right), \quad[E,F] = K, \quad [E^a,F] = 0 \nn \\
[E'^a,F] &= E_a, \quad [E'^a,F_b] = - \delta^a_b E, \quad [E'^a,F'_b] = [E^a,F_b] \nn \\
[E^{ab},F_c] &= 4 E^{[a} \delta^{b]}_c, \quad [E^{ab},F_{cd}]= 16 K\indices{^{[a}_{[c}}\delta^{b]}_{d]} - 4 \delta^{[a}_c \delta^{b]}_d \left(\sum_{e=1}^5 K\indices{^e_e} + K\right)
\end{align}
Some of of the commutators between raising and lowering operators are automatically zero because $l$ and $l'$ must be of the same sign. For example, $[E^a,F]$ and $[E^{ab},F'_b]$ would be on level $(1,-1)$, $[E^{ab},F]$ on level $(2,-1)$ and so are necessarily zero.

\subsection{Scalar products}

To conclude, we give the scalar products between the generators of Table {\bf 1} that we shall need below.  These are
\begin{align}
& (K\indices{^a_b} | K\indices{^c_d}) = \delta^a_d \delta^c_b - \delta^a_b \delta^c_d , \; \; (K | K) = 4 \nn \\
& (E|F) = 2 , \; \;  (E^a|F_b) = 2 \delta^a_b , \; \;  (E'^a|F'_b) = 2 \delta^a_b \nn \\
& (E^{ab}|F_{cd}) = 4.2! \,\delta^{[a}_c \delta^{b]}_d , \; \; 
(E'^{ab}|F'_{cd}) = 8.2! \,\delta^{[a}_c \delta^{b]}_d , \; \; 
(E''^{ab}|F''_{cd}) = 16.2! \,\delta^{[a}_c \delta^{b]}_d \nn \\
&(E^{abc}|F_{def}) = 8.3! \,\delta^{[a}_d \delta^b_e \delta^{c]}_f , \; \; 
(E'^{abc}|F'_{def}) = 8.3! \,\delta^{[a}_d \delta^b_e \delta^{c]}_f , \nn \\
&(E^{abcd}|F_{efgh}) = 8.4! \,\delta^{[a}_e \delta^b_f \delta^c_g \delta^{d]}_h
\end{align}
To derive the scalar products, one proceeds recursively using the invariance property.

\section{Sigma model  (up to level $(4,1)$)}
\setcounter{equation}{0}
\label{Sec:Sigma}

\subsection{Lagrangian}
To derive the Lagrangian for the coset space $F_4^{++}/K(F_4^{++})$, we follow the standard method. 

We recall that  $K(F_4^{++})$ is the subalgebra invariant under the Chevalley involution. In the case of the split form of finite-dimensional algebras, this subalgebra is the maximal compact subalgebra. 

By a $K(F_4^{++})$-transformation, one can always map an element of $F_4^{++}$ on the non negative part of the algebra. We will impose this condition except for the gravitational subalgebra $A_4$, for which we shall keep the negative components.  In that (partial) ``Borel'' (or ``triangular'') gauge, the coset representative is thus chosen to be
\be V(t) = H(t)T(t) = \exp\left[ h\indices{_a^b}(t) K\indices{^a_b} + \frac{1}{2}\varphi(t) K\right] \exp \left[ A(t) \right]  \ee
where
\begin{align}
A(t) = \frac{1}{\sqrt 2} \bigg( &\psi(t) E + A_a(t) E^a + B_a(t) E'^a + \frac{1}{2!} A_{ab}(t) E^{ab} + \frac{1}{2!} B_{ab}(t) E'^{ab}  \nn \\
&+ \frac{1}{2!} C_{ab}(t) E''^{ab} + \frac{1}{3!} A_{abc}(t) E^{abc} + \frac{1}{3!} B_{abc}(t) E'^{abc} + \frac{1}{4!} A_{abcd}(t) E^{abcd} \bigg)
\end{align}
This expression defines the various fields of the theory up to level $(4,1)$.  Truncation up to that level is consistent for the same reasons as given in \cite{Damour:2002cu,Damour:2002et} for $E_{10}$. There are only antisymmetric fields (``$p$-forms'') in the $A$-factor.  Anticipating the comparison with chiral supergravity, we shall call $e\indices{_a^b} \equiv \left(e^h\right)\indices{_a^ b}$ the ``vielbein'', $\varphi$ the ``dilaton'' and $\psi$ the ``axion''. The dual graviton appears at level $(4,2)$ and will be discussed below.
In terms of
\be P(t) = \frac{1}{2} \left[ \partial V(t) V^{-1}(t) - \tau\left( \partial V(t) V^{-1}(t) \right)\right] \ee
($\partial$ is the time derivative), the Lagrangian is then
\be n \mathcal{L} = (P | P ), \ee
where $(\cdot | \cdot)$ is the invariant bilinear form on $F_4^{++}$ given above and where $n$ is the (rescaled) lapse that implements the Hamiltonian constraint and ensures that the motion is a lightlike geodesic \cite{Damour:2002cu,Damour:2002et}.

We have
\[ \partial V V^{-1} = \partial H H^{-1} + H (\partial T T^{-1}) H^{-1} \]
The term $(\partial H) H^{-1}$ differs from the usual purely gravitational contribution by the dilaton term
\[  \frac{1}{2} (\partial\varphi) K. \]  As for $(\partial T) T^{-1}$, a direct but somewhat tedious computation yields
\begin{align}
(\partial T) T^{-1} = \frac{1}{\sqrt 2} \bigg( \D \psi E &+ \D A_a E^a + \D B_a E'^a + \frac{1}{2!} \D A_{ab} E^{ab} + \frac{1}{2!} \D B_{ab} E'^{ab} + \frac{1}{2!} \D C_{ab} E''^{ab} \nn \\
&+ \frac{1}{3!} \D A_{abc} E^{abc} + \frac{1}{3!} \D B_{abc} E'^{abc} + \frac{1}{4!} \D A_{abcd} E^{abcd} \bigg)
\end{align}
where the covariant derivatives are given by
\begin{align}
& \D\psi = \partial \psi , \; \;  
\D A_a = \partial A_a , \; \; 
\D B_a = \partial B_a + \frac{1}{2 \sqrt 2} ( A_a \partial \psi - \psi \partial A_a) 
\end{align}
for the axion $\psi$ and the one-forms $A_a$, $B_a$, by
\begin{subequations}
\begin{align}
\D A_{ab} &= \partial A_{ab} + \frac{1}{\sqrt 2} A_{[a}\partial A_{b]} \\
\D B_{ab} &= \partial B_{ab} + \frac{1}{2 \sqrt 2} \left(  - \psi \partial A_{ab} + A_{[a} \partial B_{b]} + B_{[a} \partial A_{b]} + A_{ab} \partial \psi \right) - \frac{1}{4} \psi A_{[a} \partial A_{b]} \\
\D C_{ab} &= \partial C_{ab} + \frac{1}{2\sqrt 2} \left( - \psi \partial B_{ab} + B_{[a} \partial B_{b]} + B_{ab} \partial \psi \right) \nn \\
&\qquad + \frac{1}{12} \left( \psi^2 \partial A_{ab} - \psi A_{[a} \partial B_{b]} - 2 \psi B_{[a} \partial A_{b]} - \psi \partial \psi A_{ab} + B_{[a} A_{b]} \partial \psi   \right) \nn \\
&\qquad + \frac{1}{16\sqrt 2} \psi^2 A_{[a} \partial A_{b]}
\end{align}
\end{subequations}
for the two-forms $A_{ab}$, $B_{ab}$, $C_{ab}$,  by
\bea
\D A_{abc} &&= \partial A_{abc} + \frac{3}{2 \sqrt 2} \left( -A_{[a} \partial B_{bc]} + B_{[a} \partial A_{bc]} - A_{[ab} \partial B_{c]} + B_{[ab} \partial A_{c]} \right) \nn \\
&&\qquad + \frac{1}{4}\left( \psi A_{[a} \partial A_{bc]} - 3 A_{[a} B_b \partial A_{c]} - 2 A_{[a} A_{bc]} \partial \psi + \psi A_{[ab} \partial A_{c]} \right) 
\eea
and
\begin{align}
\D B_{abc} &= \partial B_{abc} + \frac{1}{2\sqrt 2} \left( - \psi \partial A_{abc} - 6 A_{[a} \partial C_{bc]} + 3 B_{[a} \partial B_{bc]} - 3 B_{[ab} \partial B_{c]} \right. \nn \\
&\hphantom{= \partial B_{abc} + \frac{1}{2\sqrt 2} (}\quad \left. + 6 C_{[ab} \partial A_{c]} + A_{abc} \partial\psi \right) \nn \\
&\qquad + \frac{1}{4} \left( 3 \psi A_{[a} \partial B_{bc]} - 2 \psi B_{[a} \partial A_{bc]} + \psi A_{[ab} \partial B_{c]} - 3 A_{[a} B_b \partial B_{c]} \right. \nn \\
&\hphantom{\qquad + \frac{1}{4} (}\quad \left.- 3 A_{[a} B_{bc]} \partial \psi + B_{[a} A_{bc]} \partial \psi \right) \nn \\
&\qquad + \frac{1}{16 \sqrt 2} \left( -3 \psi^2 A_{[a} \partial A_{bc]} +10 \psi A_{[a} B_b \partial A_{c]} + 4 \psi \partial \psi A_{[a} A_{bc]} - \psi^2 A_{[ab} \partial A_{c]} \right) 
\end{align}
for the three-forms $A_{abc}$,  $B_{abc}$, and by
\begin{align}
\D A_{abcd} &= \partial A_{abcd} + \sqrt 2 \left( A_{[a} \partial A_{bcd]} - 3 A_{[ab} \partial B_{cd]} + 3 B_{[ab} \partial A_{cd]} + A_{[abc} \partial A_{d]} \right) \nn \\
&\qquad + A_{[a} B_b \partial A_{cd]} - 2 A_{[a} A_{bc} \partial B_{d]} + 3 A_{[a} B_{bc} \partial A_{d]} + \psi A_{[ab} \partial A_{cd]} \nn \\
&\qquad - A_{[ab} B_c \partial A_{d]} - A_{[ab} A_{cd]} \partial \psi \nn \\
&\qquad + \frac{1}{\sqrt 2} \psi A_{[a} A_{bc} \partial A_{d]}
\end{align}
for the four-form $A_{abcd}$.

To compute $H (\partial T T^{-1}) H^{-1}$ from $e^A B e^{-A} = e^{\ad_A} B$, we use formula (\ref{actionofK}).  The Lagrangian is then found to be
\bea
n \mathcal{L}_{F_4^{++}/K(F_4^{++})} &&= \frac{1}{4} \left( g^{ac}g^{bd} - g^{ab}g^{cd} \right) \partial g_{ab} \partial g_{cd} + (\partial\varphi)^2  \nn \\
&&\qquad + \frac{1}{2}e^{2\varphi}(\D\psi)^2 + \frac{1}{2}e^{-\varphi}\D A_a \D A^a + \frac{1}{2}e^{\varphi}\D B_a \D B^a  \nn \\
&&\qquad + \frac{1}{2}e^{-2\varphi}\D A_{ab} \D A^{ab} + \D B_{ab} \D B^{ab} + 2 e^{2\varphi}\D C_{ab} \D C^{ab}  \nn \\
&&\qquad + \frac{1}{3}e^{-\varphi}\D A_{abc} \D A^{abc} + \frac{1}{3}e^{\varphi}\D B_{abc} \D B^{abc} \nn \\
&& \qquad + \frac{1}{12}e^{-2\varphi}\D A_{abcd} \D A^{abcd} \label{LagrangianSigma}
\eea
where the metric $g_{ab}$ is related to the vielbein through
\be
 g_{ab} = \sum_c e\indices{_a^c} e\indices{_b^c} .
\ee

\subsection{Equations of motion}
The equations of motion that follow from the Lagrangian are, with the gauge choice $n=1$:

\vspace{.2cm}
\noindent
(i) 4-form: \be
\partial \!\left( e^{-2\varphi} \D A_{abcd} \right) = 0 
\ee
(ii) 3-forms:
\be \partial \!\left( e^{\varphi} \D B_{abc} \right) = 0 , \; \; 
\partial \!\left( e^{-\varphi} \D A_{abc} \right) = \frac{1}{\sqrt 2} e^\varphi \D B_{abc} \D \psi + \frac{1}{\sqrt 2} e^{-2 \varphi} \D A_{abcd} \D A^d 
\ee
(iii) 2-forms
\begin{align}
&\partial \!\left( e^{2\varphi} \D C_{ab} \right) = \frac{1}{\sqrt 2} e^\varphi \D B_{abc} \D A^c \\
&\partial \D B_{ab} = \frac{1}{\sqrt 2}e^{-2 \varphi} \D A_{abcd} \D A^{cd} - \frac{1}{\sqrt 2}e^{\varphi} \D B_{abc} \D B^{c} + \frac{1}{\sqrt 2}e^{-\varphi} \D A_{abc} \D A^{c} \nn \\
&\hphantom{\partial \D B_{ab} =}  + \sqrt{2} e^{2\varphi} \D C_{ab} \D \psi \\
&\partial \!\left( e^{-2\varphi} \D A_{ab} \right) = -\sqrt{2}e^{-2 \varphi} \D A_{abcd} \D B^{cd} - \sqrt{2} e^{- \varphi} \D A_{abc} \D B^{c} + \sqrt 2 \D B_{ab} \D \psi 
\end{align}
(iv) 1-forms
\begin{align}
&\partial \!\left( e^\varphi \D B_a \right) = \sqrt{2} e^{\varphi} \D B_{abc} \D B^{cd} + \sqrt{2} e^{-\varphi} \D A_{abc} \D A^{bc} + 2\sqrt{2} e^{2\varphi} \D C_{ab} \D B^b \nn \\
&\hphantom{\partial \!\left( e^\varphi \D B_a \right) =} + \sqrt{2} \D B_{ab} \D A^b  \\
&\partial \!\left( e^{-\varphi} \D A_a \right) = \frac{\sqrt 2}{3} e^{-2 \varphi} \D A_{abcd} \D A^{bcd} - 2\sqrt{2} e^{\varphi} \D B_{abc} \D C^{bc} - \sqrt{2} e^{-\varphi} \D A_{abc} \D B^{bc} \nn \\
&\hphantom{\partial \!\left( e^{-\varphi} \D A_a \right) =} + \sqrt 2 \D B_{ab} \D B^b + \sqrt 2 e^{-2\varphi} \D A_{ab} \D A^b + \frac{1}{\sqrt 2} e^{\varphi} \D B_a \D \psi 
\end{align}
(v) Axion
\begin{align}
&\partial \!\left( e^{2\varphi} \D \psi \right) = - \frac{\sqrt 2}{3} e^{\varphi} \D B_{abc} \D A^{abc} - 2 \sqrt 2 e^{2\varphi} \D C_{ab} \D B^{ab} - \sqrt 2 \D B_{ab} \D A^{ab} \nn \\
&\hphantom{\partial \!\left( e^{2\varphi} \D \psi \right) =} - \frac{1}{\sqrt 2} e^\varphi \D B_a \D A^a
\end{align}
(vi) Dilaton
\begin{align}
&\partial^2 \varphi = \frac{1}{2}e^{2\varphi}(\D\psi)^2 - \frac{1}{4}e^{-\varphi}\D A_a \D A^a + \frac{1}{4}e^{\varphi}\D B_a \D B^a - \frac{1}{2}e^{-2\varphi}\D A_{ab} \D A^{ab} \nn \\
&\hphantom{\partial^2 \varphi =} + 2 e^{2\varphi}\D C_{ab} \D C^{ab} - \frac{1}{6}e^{-\varphi}\D A_{abc} \D A^{abc} + \frac{1}{6}e^{\varphi}\D B_{abc} \D B^{abc} - \frac{1}{12}e^{-2\varphi}\D A_{abcd} \D A^{abcd} 
\end{align}
(vii) Metric
\begin{align}
&\frac{1}{2} \partial\!\left( g^{ac} \partial g_{cb} \right) = \frac{1}{4} \delta^a_b \left( \frac{1}{2}e^{-\varphi}\D A_c \D A^c + \frac{1}{2}e^{\varphi}\D B_c \D B^c \right. \nn \\
&\hphantom{\frac{1}{2} \partial\!\left( g^{ac} \partial g_{cb} \right) = \frac{1}{4} \delta^a_b (} + e^{-2\varphi}\D A_{cd} \D A^{cd} + 2 \D B_{cd} \D B^{cd} + 4 e^{2\varphi}\D C_{cd} \D C^{cd} \nn \\
&\hphantom{\frac{1}{2} \partial\!\left( g^{ac} \partial g_{cb} \right) = \frac{1}{4} \delta^a_b (} \left. + e^{-\varphi}\D A_{cde} \D A^{cde} + e^{\varphi}\D B_{cde} \D B^{cde} + \frac{1}{3}e^{-2\varphi}\D A_{cdef} \D A^{cdef} \right) \nn \\
&\hphantom{\frac{1}{2} \partial\!\left( g^{ac} \partial g_{cb} \right) =} -\frac{1}{2}e^{-\varphi}\D A^a \D A_b - \frac{1}{2}e^{\varphi}\D B^a \D B_b \nn \\
&\hphantom{\frac{1}{2} \partial\!\left( g^{ac} \partial g_{cb} \right) =} - e^{-2\varphi}\D A^{ac} \D A_{bc} - 2 \D B^{ac} \D B_{bc} - 4 e^{2\varphi}\D C^{ac} \D C_{bc} 
\nn \\
&\hphantom{\frac{1}{2} \partial\!\left( g^{ac} \partial g_{cb} \right) =} - e^{-\varphi}\D A^{acd} \D A_{bcd} - e^{\varphi}\D B^{acd} \D B_{bcd} - \frac{1}{3}e^{-2\varphi}\D A^{acde} \D A_{bcde}
\end{align}

Finally, the Hamiltonian constraint, obtained by extremizing the action with respect to $n$, reads
\bea
0 &=& \frac{1}{4} \left( g^{ac}g^{bd} - g^{ab}g^{cd} \right) \partial g_{ab} \partial g_{cd} + (\partial\varphi)^2  \nn \\
&&+ \frac{1}{2}e^{2\varphi}(\D\psi)^2 + \frac{1}{2}e^{-\varphi}\D A_a \D A^a + \frac{1}{2}e^{\varphi}\D B_a \D B^a \nn \\
&& + \frac{1}{2}e^{-2\varphi}\D A_{ab} \D A^{ab} + \D B_{ab} \D B^{ab} + 2 e^{2\varphi}\D C_{ab} \D C^{ab} \nn \\
&& + \frac{1}{3}e^{-\varphi}\D A_{abc} \D A^{abc} + \frac{1}{3}e^{\varphi}\D B_{abc} \D B^{abc} + \frac{1}{12}e^{-2\varphi}\D A_{abcd} \D A^{abcd}
\eea

\section{Correspondence with the gravitational model up to level $(4,1)$}
\setcounter{equation}{0}
\label{Sec:Comparison}

\subsection{Homogeneous fields}
We follow again the approach of \cite{Damour:2002cu}.  The comparison between the supergravity field equations and the sigma model equations should be thought of as being carried out in some generalized (and still to be completely specified) form of spatial gradient expansion. 

At lowest order in that expansion, the fields on the supergravity side are  taken to be spatially homogeneous, i.e., to depend only on time,
\begin{subequations}
\label{Homo1}
\begin{align}
ds^2 &= - \mathrm{g}(t) dt^2 + \mathrm{g}_{ab}(t) dx^a dx^b \quad (\mathrm{g} = \det ( \mathrm{g}_{ab} ))\\
\phi &= \phi(t) \\
\partial_\mu \chi &= \partial_\mu \chi (t), \quad F^{\pm}_{\mu\nu} = F^{\pm}_{\mu\nu}(t) \\
H_{\mu\nu\rho}&=H_{\mu\nu\rho}(t), \quad G_{\mu\nu\rho}=G_{\mu\nu\rho}(t) .
\end{align}
\end{subequations}
We also make the gauge choice $N=\sqrt{\mathrm{g}}$ for the lapse (corresponding to $n=1$ on the sigma model side) and $N^k=0$ for the shift.
Note that we allow both electric and magnetic components for the axion and the $p$-form fields.  This means that we go beyond the assumption of spatially homogeneous potentials, which would yield only non-vanishing electric fields.

On the sigma model side, we truncate the equations by retaining fields only up to level $(4,1)$, as we already did above. Had we kept only the electric fields on the supergravity side,  we should truncate the sigma model up to level $(2,0)$ (or $(2,1)$ depending on how one views the field strength of the chiral $2$-form).  Keeping the magnetic fields enables one to test the conjecture at higher levels.

\subsection{Dictionary}
Given these truncations, one finds that the equations of motion of chiral supergravity and of the coset model perfectly match if we make the identifications
\begin{subequations}
\begin{align}
\mathrm{g}_{ab} &\longleftrightarrow g_{ab} \\
\phi &\longleftrightarrow \varphi \\
\dot\chi &\longleftrightarrow \D\psi \\
F^{-}_{0a} &\longleftrightarrow -\D A_a \\
F^{+}_{0a} &\longleftrightarrow +\D B_a \\
H_{0ab} &\longleftrightarrow - \sqrt 2 \,\D A_{ab} \\
G\indices{_{0ab}} = (\star G)_{0ab} = \frac{\mathrm g}{3!} \varepsilon_{0abklm} G^{klm}  &\longleftrightarrow -\sqrt 2 \,\D B_{ab}  \label{SigmaDuality}\\
(\star H)_{0ab} = \frac{\mathrm{g}}{3!} \varepsilon\indices{_{0abklm}}H^{klm} &\longleftrightarrow 2\sqrt 2 e^{2\varphi} \D C_{ab} \displaybreak \\
(\star F^+)_{0abc} = \frac{\mathrm{g}}{2} \varepsilon\indices{_{0abckl}}F^{+kl} &\longleftrightarrow -2 e^{-\varphi} \D A_{abc} \\
(\star F^-)_{0abc} = \frac{\mathrm{g}}{2} \varepsilon\indices{_{0abckl}}F^{-kl} &\longleftrightarrow -2 e^{\varphi} \D B_{abc} \\
(\star d\chi)_{0abcd} = \mathrm{g} \varepsilon\indices{_{0abcdk}}\partial^k \chi &\longleftrightarrow 2 e^{-2\varphi} \D A_{abcd}
\end{align}
\end{subequations}
Not only do the dynamical equation of motion match, but also the Hamiltonian constraint does. 

In particular, all the Chern-Simons couplings between the $p$-forms are exactly reproduced by the sigma model Lagrangian.   This is remarkable because these couplings are derived, in the standard approach, by using supersymmetry.  This is one more instance of the intriguing connection between the hidden symmetry and supersymmetry, which seem to be independepent concepts but yet give identical predictions.

It is quite appealing that the self-duality condition on the field strength of the $2$-form $C$ is naturally incorporated in the sigma model.  How does this arise? For each non-chiral $p$-form, the standard $p$-form potential and its dual potential appear simultaneously in the field content of the sigma model.  In the geodesic equations of motion, the electric fields of both  occur, and the electric field of the dual potential is identified with the magnetic field of the standard potential. This is a familiar fact which actually holds already for the $E_{10}$ model. For the chiral $2$-form, however, there is only {\em one} potential, so that one must identify its electric and magnetic fields in the dictionary.  This is what was done in (\ref{SigmaDuality}).

\subsection{Beyond level $(4,1)$}

Except for $\phi$ and $g$, whose duals appear at level $(4,2)$, the duals of all the supergravity fields are present in the truncation up to level $(4,1)$.  One can go beyond level $(4,1)$ by including more spatial gradients on the supergravity side.  One way to proceed is to replace the abelian homogeneity group leading to the form (\ref{Homo1}) of the fields (``Bianchi type I'') by a non-abelian group along the lines of \cite{Demaret:1985js}, which allows non-vanishing spatial gradients in a controlled way.   Alternatively, one may consider the next terms in the gradient expansion of the supergravity field around an arbitrarily chosen spatial point.  Either way, one would find that the matching extends up up to level $(4,2)$ and $(4,3)$, but this matching requires some well-chosen gauge conditions in order for one to be able to consistently identify the $(3,1)$-mixed Young field with the dual gravity (through the spatial anholonomy) and in that sense may be argued to be less understood. Even though we have not checked it explicitely, we expect the details to work in the same way as for the $E_{10}$ model \cite{Damour:2002cu}.

Similarly, while the Hamiltonian constraints on both sides of the correspondence nicely match, the other supergravity constraints must be implemented on the sigma model side.  This raises interesting questions which have been explored in the important work  \cite{Damour:2007dt,Damour:2009ww,Kleinschmidt:2014uwa}, but which still needs further study. These other supergravity constraints are the momentum constraints and the various Gauss' laws.

\section{$V$-duality and Borcherds superalgebra} 
\setcounter{equation}{0}
\label{Sec:Borcherds}

\subsection{Cartan matrix}
The structure of the equations discussed above is very similar to that encountered in type IIB supergravity in $D=10$ dimensions where there is a chiral 4-form, the curvature of which is self-dual. This self-duality condition is also properly incorporated in the sigma model formulation \cite{Kleinschmidt:2004rg}. The $p$-form content is, however, different and this can best be discussed in terms of the underlying Borcherds algebras \cite{HenryLabordere:2002dk,Henneaux:2010ys}.  

In the $F_4^{++}$ spectrum, all forms can be constructed by successive commutation (and antisymmetrization in the indices) of the $E^a$ and $E$ generators. From these, we construct the raising generators of a Borcherds superalgebra as 
\begin{align}
e_1 &= E^a \theta_a \nn \\
e_2 &= E,
\end{align}
where the $\theta_a$'s are a basis of 1-forms that automatically implement the antisymmetrization.  Thus $e_1$ is a fermionic (odd) generator while $e_2$ is bosonic (even).  In  $F_4^{++}$, the index $a$ takes values from $1$ to $5$, but we shall lift that condition from now on and not specify the dimension of space so as to investigate forms of higher rank. 
From the $F_4^{++}$ commutation relations, we find the only Serre relation
\be (\ad_{e_2})^2 e_1 = [e_2,[e_2,e_1]] = 0. \ee

We now show how to extend the generators $\{e_1, e_2\}$  to the Chevalley-Serre generators of a Borcherds superalgebra. The Cartan subalgebra is spanned by the $\mathfrak{gl}(5,\mathbb{R})$ trace $ H \equiv \sum_a K\indices{^a_a}$ and the generator $K$. They have the following commutation relations with the $e_i$'s:
\begin{align}
&[H,e_1] = e_1, \quad [H,e_2] = 0 \nn \\
&[K, e_1] = - e_1, \quad [K, e_2] = 2 e_2
\end{align}
If we take the linear combination
\begin{align}
h_1 &= \left( k-\frac12 \right) H - \frac{1}{2} K \nn \\
h_2 &= K,
\end{align}
where $k$ is an arbitrary constant satisfying $k < 0$, then the $h_i$, $e_i$ and $f_i=-\tau(e_i)$ generate a Borcherds superalgebra with Cartan matrix
\be A = \begin{pmatrix} k & -1 \\ -1 & 2 \end{pmatrix}. \label{CB0} \ee
 The fact that $k<0$ implies that there is no condition on the graded commutator (anticommutator) $[e_1,e_1]$ (recall that the first root is fermionic).  If $k$ were to vanish, one would have the Serre relation $[e_1,e_1] =0$, but this relation does not hold in $F_4^{++}$.

The choice of the constant $k$ does not affect the $p$-form spectrum; however, there is a natural choice, $k=-2$, to be explained below.

\subsection{$p$-form spectrum}

The reason that the exact value of $k$ does not affect the $p$-form spectrum is that the only Serre relation is $[e_2, [e_2, e_1]] = 0$ not matter what $k$ is (provided $ k<0$). 
Along with the (graded) Jacobi identity, this Serre relation suffices to determine the $p$-form spectrum by taking successive graded commutators, since only the relations between the raising operators $e_i$ are needed.  Each independent graded commutator containing $l$ times $e_1$ and $l'$ times $e_2$ correspond to a $l$-form in the spectrum at level $(l,l')$. The number of such forms is written $m_{l,l'}$ in the table below.
The result is
\begin{center}
{\bf Table 2: } $p$-form spectrum

\begin{tabular}{c|c}
level $(l,l')$ & multiplicity $m_{l,l'}$ \\ \hline
$(0,1)$ & 1 \\ \hline
$(1,0)$ & 1 \\
$(1,1)$ & 1 \\ \hline 
$(2,0)$ & 1 \\
$(2,1)$ & 1 \\
$(2,2)$ & 1 \\ \hline
$(3,1)$ & 1 \\
$(3,2)$ & 1 \\ \hline
$(4,1)$ & 1 \\
$(4,2)$ & 1 \\
$(4,3)$ & 1 \\ \hline
$(5,1)$ & 1 \\
$(5,2)$ & 2 \\
$(5,3)$ & 2 \\
$(5,4)$ & 1 \\ \hline
$(6,1)$ & 1 \\
$(6,2)$ & 3 \\
$(6,3)$ & 3 \\
$(6,4)$ & 3 \\
$(6,5)$ & 1
\end{tabular}
\end{center}
For instance, the generator at level $(1,1)$ is $[e_1, e_2]$, that at level $(2,0)$ is $[e_1, e_1]$ etc.  

Note that the table agrees with the data given by the level decomposition of $F_4^{+++}$ (see A.6 of \cite{Kleinschmidt:2003mf} and \cite{Kleinschmidt:2003jf,Riccioni:2007hm}). In particular, we can see in those tables that all $p$-forms with $p \leq 5$ indeed belong to $F_4^{++}$ as expected from the truncation $a=1, 2, 3, 4, 5$ for $\theta_a$, while the 6-forms are specific to $F_4^{+++}$, and indeed do have a non-vanishing coefficient along the very extended root. Moreover, we also see here that the generators fall into representations of $\mathfrak{sl}(2)$: at level $l=5$, we have the representations {\bf 4} and {\bf 2}, and at level $l=6$, we have the {\bf 5} and two times the {\bf 3}, in agreement with \cite{Riccioni:2007hm}.

Instead of constructing the generators at higher levels in a pedestrian fashion, which is direct at low levels, one can apply the denominator formula (see e.g. \cite{DenFor}). In our case, this  formula reads
\be \frac{\prod_{\alpha \in \Delta_0^+}(1-e^{-\alpha})^{m(\alpha)}}{\prod_{\beta \in \Delta_1^+}(1+e^{-\beta})^{m(\beta)}}\ = 1 - e^{-\alpha_1} - e^{-\alpha_2} + e^{-\alpha_1 - 2 \alpha_2}, \ee
where $\Delta_0^+$ (resp. $\Delta_1^+$) is the set of positive even (resp. odd) roots, $m(\alpha)$ is the multiplicity of the root $\alpha$ (if $\alpha$ is not a root, then $m(\alpha)=0$), $\alpha_1$ and $\alpha_2$ are the simple roots of our algebra ($\alpha_1$ is odd, $\alpha_2$ is even). This formula allows us to find all the desired multiplicities $m_{l,l'}=m(l\alpha_1 + l'\alpha_2)$.

To make it more useable, we note that the positive even roots are all of the form $\alpha=2k \alpha_1 + l' \alpha_2$ and that the odd ones are of the form $\beta= (2k+1) \alpha_1 + l' \alpha_2$, where $k$ and $l'$ are nonnegative integers. Defining the formal variables $x=e^{-\alpha_1}$ and $y=e^{-\alpha_2}$, the denominator identity takes the form
\be \prod_{k,l'=0}^\infty \frac{(1-x^{2k}y^{l'})^{m_{2k,l'}}}{(1+x^{2k+1}y^{l'})^{m_{2k+1,l'}}} = 1 -x -y + xy^2 \ee
The expansion of the left-hand side in a power series allows us to read off the numbers $m_{l,l'}$. This gives the results of the table.

Explicitely, up to $l=2$:
\begin{itemize}
\item $l=0$:
We need only keep the terms that contain no $x$: this gives
\[ \prod_{l'=0}^\infty (1-y^{l'})^{m_{0l'}} = 1-y \]
from which we see that $m_{01}=1$ while $m_{0l'}=0$ for all $l' \geq 2$.
\item $l=1$:
\[ \prod_{l'=0}^\infty (1-y)(1+xy^{l'})^{-m_{1l'}} = 1-y -x -xy^2 \]
We see that $m_{1l'}=0$ for all $l'\geq 3$, since there are no terms of the form $xy^{l'}$ with $l'\geq 3$ on the right hand side.
Forgetting all terms of order $x^2$ and higher, we expand the left hand side as
\begin{align*}
(1-y)&(1+x)^{-m_{10}}(1+xy)^{-m_{11}}(1+xy)^{-m_{12}} \\
&= (1-y)(1-m_{10}x)(1-m_{11}xy)(1-m_{12}xy^2) \\
&=1-y-m_{10}x+(m_{10}-m_{11})xy + (m_{11}-m_{12}) xy^2 + m_{12}xy^3
\end{align*}
and we read off the numbers $m_{10}=1$, $m_{11}=1$, $m_{12}=0$.
\item $l=2$:
\[ \prod_{l'=0}^\infty (1-y)(1+x)^{-1}(1+xy)^{-1}(1-x^2y^{l'})^{m_{2l'}} = 1-y-x-xy^2 \]
Up to order $x^2$, the first three factors are
\begin{align*}
(1-y)(1+x)^{-1}(1+xy)^{-1}&=(1-y)(1-x+x^2)(1-xy+x^2y^2)\\
&=1-y-x+xy^2+x^2-x^2y^3
\end{align*}
and we have, keeping only the $x^2$ terms,
\begin{align*}
\prod_{l'=0}^\infty &(1-y-x+xy^2+x^2-x^2y^3)(1-m_{2l'} x^2y^{l'}) \\
&= x^2-x^2y^3 + \sum_{l'=0}^\infty (-m_{2l'}x^2 y^{l'} + m_{2l'}x^2y^{l'+1}).
\end{align*}
We see that $m_{2l'}=0$ for all $l' \geq 3$.
As there are no $x^2$ terms on the right hand side of the denominator formula, this gives
\[ (1-m_{20})x^2 + (-m_{21}+m_{20})x^2y + (-m_{22}+m_{21})x^2y^2 + (-1-m_{23}+m_{22})x^2y^3 = 0 \]
so that $m_{20}=m_{21}=m_{22}=1$ and $m_{23}=0$.
\end{itemize}
This can be continued up to arbitrary $l$, each time using the information gained at smaller $l$.

\subsection{Comparing with type IIB}

To compare the Borcherds superalgebra describing the $V$-duality of chiral supergravity in six dimensions with the Borcherds algebra describing the $V$-duality of type IIB supegravity in ten dimensions, we first need to determine $k$. 

To that end, we follow the method of \cite{Kleinschmidt:2013em}, which consists in starting from the Borcherds superalgebra in lower dimensions where there is no ambiguity and oxidizing according to a well-defined procedure.

We start in three spacetime dimensions, where the symmetry is $F_4$  with simple roots denoted $\beta_i$ ($i=1,2,3,4$). Their matrix of scalar products is
\be \beta_i \cdot \beta_j = \begin{pmatrix}
2 & -1 & 0 & 0 \\ 
-1 & 2 & -1 & 0 \\ 
0 & -1 & 1 & -1/2 \\ 
0 & 0 & -1/2 & 1
\end{pmatrix}. \ee
The relevant Borcherds superalgebra is obtained by adding a null fermionic root $\gamma_0$, connected only to $\beta_1$ in the Dynkin diagram of $F_4$, i.e. $\gamma_0 \cdot \gamma_0 = 0$, $\gamma_0 \cdot \beta_1 = -1$, $\gamma_0 \cdot \beta_i = 0$ for $i=2,3,4$ \cite{Henneaux:2010ys}. This gives the Borcherds symmetry in 3 dimensions, which we can oxidize up to 6 dimensions. We get successively:\begin{itemize}
\item $D=4$: the roots of the algebra are $\gamma_1=\gamma_0 + \beta_1$, $\beta_2$, $\beta_3$ and $\beta_4$. We have $\gamma_1 \cdot \gamma_1 = 0$, $\gamma_1 \cdot \beta_2 = -1$, $\gamma_1 \cdot \beta_i = 0$ for $i=3,4$.
\item $D=5$: the roots are $\gamma_2=\gamma_1 + \beta_2$, $\beta_3$ and $\beta_4$. We have $\gamma_2 \cdot \gamma_2 = 0$, $\gamma_2 \cdot \beta_3 = -1$, $\gamma_2 \cdot \beta_4 = 0$.
\item $D=6$: the roots are $\gamma_3=\gamma_2 + \beta_3$ and $\beta_4$. We have $\gamma_3 \cdot \gamma_3 = -1$ and $\gamma_3 \cdot \beta_4 = -1/2$.
\end{itemize}
We end up with a Borcherds superalgebra that contains a fermionic root $\gamma_3$ and a bosonic root $\beta_4$. Their matrix of scalar products is
\be \begin{pmatrix} -1 & -1/2 \\ -1/2 & 1 \end{pmatrix}. \ee
To put this matrix in the form (\ref{CB0}),  we make the rescaling $h_i = 2 \bar{h}_i$ to get the Cartan matrix
\be \begin{pmatrix} -2 & -1 \\ -1 & 2 \end{pmatrix} \ee
which fixes $k=-2$.  This is the Cartan matrix of a Borcherds algebra with generators $\bar{h}_i$, $e_i$ and $f_i$, which is isomorphic to our algebra. 

It turns out that this Cartan matrix is very similar to the Cartan matrix for type IIB obtained by following the same procedure \cite{Kleinschmidt:2013em} in the sense that both contain one timelike simple root and one spacelike simple root. A difference lies in the grading of the generators.   In the first case there is one fermionic generator (one-form) and one bosonic generator (zero-form) so it is a genuine superalgebra, while in the second case, both generators are bosonic (a two-form and a zero-form).  The tight connection between the two theories has of course already been noticed before.  We see here that it also appears when one considers the $V$-dualities.

\section{Concluding remarks \label{sec:Concluding-remarks}}
\setcounter{equation}{0}

In this paper, we have investigated the equations for the geodesic motion on the coset space $F_4^{++} /K(F_4^{++})$ and shown their equivalence with the equations of motion of six-dimensional chiral supergravity with two vector multiplets and two tensor multiplets, up to the level where the matching starts being less understood. While this agreement was expected from existing experience with other supergravity models, it was interesting to see how the self-duality condition on the field strength of the chiral two-form emerged on the coset model side. The way it is implemented can be summarized as follows.  Only ``electric fields'' (covariant time derivatives of the sigma model fields) appear in the $(1+0)$-sigma model formulation since there is no room for explicit spatial derivatives. Non-chiral forms are described by two potentials, namely, their standard potential and its dual.  One recovers the magnetic fields as the electric fields of the duals.  For the chiral form, there is, however, only one potential.  The magnetic field must then be set equal to the electric field in the dictionary.

The same phenomenon had been described earlier in the context of type IIB supergravity in ten dimensions \cite{Kleinschmidt:2004rg}.  This motivated us to compare the two models through their $p$-form spectrum, encoded in a Borcherds superalgebra structure.  We have compared the corresponding Cartan matrices and found rather close connections between the two $V$-duality algebras. 

Although the self-duality condition on the field strength of the chiral $2$-form is correctly accounted for in the sigma model, it should be noted, however, that the equations of motion of the sigma model are of second order in the time derivatives.  One does not get the self-duality condition as an equation of motion but rather as a translation rule in the dictionary that connects the sigma model variables with the supergravity fields. This raises the possibility that the sigma model Lagrangian may not provide the final word on the question of exhibiting explicitly the $F_4^{++}$ symmetry of the (possibly extended) supergravity model.

Finally, it remains a rather mysterious fact that the hidden symmetry and supersymmetry, although a priori unconnected, yield identical predictions on the structure of the Lagrangian (spectrum, coefficients of Chern-Simons terms).  To shed light on this ill-understood issue, it would be of interest to include the fermions and discuss how supersymmetry is realized in the sigma model.  It is planned to return to this problem.

\noindent \acknowledgments  We are grateful to Axel Kleinschmidt for useful discussions. We also thank Jakob Palmkvist who pointed out an error in the Cartan matrix of the Borcherds algebra of the six-dimensional model in the original version of this manuscript. M.H. thanks the Alexander von Humboldt
Foundation for a Humboldt Research Award.  Our work is partially funded by the ERC through
the ``SyDuGraM\textquotedblright{}\ Advanced Grant,  by FNRS-Belgium (convention
FRFC PDR T.1025.14 and convention IISN  4.4503.15),  by the ``Communaut\'e Fran\c{c}aise
de Belgique\textquotedblright{}\ through the ARC program and by a donation from the Solvay family.

\end{document}